\documentclass[11pt]{article}
\usepackage[T1]{fontenc}
\usepackage{lmodern}
\usepackage{amsmath}
\usepackage[margin=1in]{geometry}
\usepackage{booktabs}
\usepackage{tabularx}
\usepackage{caption}
\usepackage[numbers,sort&compress]{natbib}
\usepackage{microtype}
\usepackage{url}
\usepackage[hidelinks]{hyperref}
\makeatletter
\g@addto@macro\UrlBreaks{\do\a\do\b\do\c\do\d\do\e\do\f\do\g\do\h\do\i\do\j\do\k\do\l\do\m\do\n\do\o\do\p\do\q\do\r\do\s\do\t\do\u\do\v\do\w\do\x\do\y\do\z\do\A\do\B\do\C\do\D\do\E\do\F\do\G\do\H\do\I\do\J\do\K\do\L\do\M\do\N\do\O\do\P\do\Q\do\R\do\S\do\T\do\U\do\V\do\W\do\X\do\Y\do\Z\do\0\do\1\do\2\do\3\do\4\do\5\do\6\do\7\do\8\do\9}
\makeatother

\title{Bounded Retained History in a Proxy Flow Replay: A Mixed Predictive Pilot and an Allocation Study}

\author{Juan S. Santillana\\\texttt{juan@jsantillana.com}}
\date{Preprint. Not peer reviewed.}

\hypersetup{
  pdfauthor={Juan S. Santillana},
  pdftitle={Bounded Retained History in a Proxy Flow Replay: A Mixed Predictive Pilot and an Allocation Study},
  pdfkeywords={bounded retention, flow replay, relational history, reconnaissance attribution},
  pdfsubject={Preprint}
}

\begin{document}

\maketitle

\begin{abstract}

We examine bounded retained history through a frozen predictive pilot and a subsequent allocation study on a curated UNSW-NB15 flow mirror. Both use an artificial release-order replay based on raw \nolinkurl{ltime} ticks, with equal-tick records querying a shared prior-group state. In the pilot, the model with added retained relational features had lower reconnaissance recall than the current-flow-plus-counters baseline (43.66\% versus 47.84\%) and lower benign false-positive rate (3.63\% versus 6.68\%) at separately frozen operating thresholds, together with lower AUROC. This mixed result did not establish a predictive advantage. A development-only audit subsequently found that capacity eviction dominated expiration in the 256-record, 300-tick FIFO state, with a median completed record lifetime of 11 ticks. A protocol specified before its single execution---but after the pilot and post-hoc mechanism audit---compared five label-blind retention arms on the already-used development trace. Its primary equal-source endpoint averaged retained/\allowbreak{}reference ratios for service-expanded source out-degree and source--destination service diversity. Source-balanced retention at 256 records exceeded FIFO-256 by a pooled endpoint difference of 0.440738 and met the prespecified materiality and subblock rule. Crucially, the event-weighted ratios for both quantities decreased relative to FIFO-256: the result is specific to the equal-source weighting objective and does not establish uniformly improved retention. The study supplies neither evidence of detection improvement nor an explanation of the pilot result. Reused data, concentrated reconnaissance support, proxy timing, and the absence of a qualified independent corpus bound the interpretation.

\end{abstract}


\section{Introduction and Research Questions}

Retained-history features describe the records that remain available to a computation. Under a capacity limit, that history can differ substantially from all events inside the nominal time horizon. A predictive comparison using retained features therefore raises two distinct questions: whether the features improve the chosen prediction task, and what history the retention mechanism actually preserves.

This report addresses those questions in their observed sequence:

\begin{enumerate}
\item \textbf{Predictive-pilot question.} Under the frozen current-flow inputs, classifier family, calibration procedure, and threshold policy, did adding retained relational features improve reconnaissance attribution relative to simple retained-history counters?
\item \textbf{Retention-allocation question.} On the already-used development trace, did a specified source-balanced policy preserve more reference \ensuremath{O_s}/\allowbreak{}\ensuremath{V_{sd}} history than global FIFO at the same 256-record capacity, under a prespecified equal-source endpoint?

\end{enumerate}
The first question received a mixed answer. The second received a positive answer under its declared endpoint, accompanied by lower event-weighted retention ratios. These findings must be reported together. Neither a retained relation nor an increase in its retention ratio establishes that it carries useful predictive information.

The allocation hypothesis was motivated by an audit performed after the pilot results were observed. Although the follow-up protocol was specified before its own execution, its data and motivation were not independent of earlier work. The report consequently presents a bounded empirical account rather than a confirmation of the original predictive hypothesis. Claims about novelty and placement relative to prior methods are limited to the verified statements in Section~\ref{sec:related}; they do not establish state of the art or a new detector.

\section{Related Work}
\label{sec:related}

\paragraph{Online scan detection.}
Bro is a stand-alone system that detects network intruders in real time by passively monitoring a link; its event handlers can update state and express site policy \cite{paxson1999bro}. TRW models accesses to local addresses as a random walk and applies sequential hypothesis testing to infer whether a remote host is scanning, targeting prompt decisions with configurable error bounds \cite{jung2004trw}. The authors report detection after few connection attempts on their traces. These systems establish online scan detection as a prior problem, but they are not reproduced baselines here: this study replays completed-flow summaries and measures retained reference history, not scan decisions.

\paragraph{Bounded summaries and retention policies.}
Count-Min is a sublinear-space data structure for approximate frequency queries over data streams \cite{cormode2005countmin}. TinyLFU uses an approximate frequency summary to decide whether a newly accessed item should replace a cache victim \cite{einziger2017tinylfu}. StreamSpot performs clustering-based anomaly detection on heterogeneous graph streams and uses a bounded-size sketched graph representation \cite{manzoor2016streamspot}. These works show that bounded state and admission or eviction choices are established design dimensions. Their objectives are frequency-error bounds, cache behavior, or graph-anomaly scores; none defines the equal-source retained/reference endpoint used here, and this study does not claim to improve any of those objectives.

\paragraph{Dataset.}
UNSW-NB15 combines recorded normal activity with synthetic attack traffic and provides 49 features and nine attack categories \cite{moustafa2015unsw}. This study uses a third-party curated flow artifact derived from that collection \cite{rdpahalavan2023unsw}. The official publication does not document the mirror's row selection, and the mirror contains fewer rows than the official full collection; Section~\ref{sec:data} reports the discrepancy rather than resolving it. The dataset is a research benchmark, not an independent deployment population.

\paragraph{Position of this study.}
The cited work supplies detectors, bounded summaries, and a benchmark. This paper introduces neither a new detector nor a new sketch. It reports a frozen predictive pilot and a post-pilot allocation measurement on a reused development trace. The related work motivates retained history as an implementation dimension; it does not establish that the measured allocation difference improves detection, and no state-of-the-art comparison is made.

\section{Data, Proxy Replay, and Evaluation Restrictions}
\label{sec:data}

\subsection{Source and provenance}

The experiments use the curated UNSW-NB15 mirror at \nolinkurl{data/external/unsw_nb15/Network-Flows/UNSW_Flow.parquet}. The audited artifact contains 2,059,415 rows. The configured and execution-recorded source SHA-256 is:

\begin{quote}\small
\nolinkurl{8129db1cef95763a45072052af30a4c954af18b48abb6176f3d8ab732eb7d274}\par
\end{quote}

The retention protocol identifies repository revision \nolinkurl{25ecbc46f36eba3f2637544977904207036ac776}. The mirror is a third-party curated artifact \cite{rdpahalavan2023unsw} derived from the UNSW-NB15 collection \cite{moustafa2015unsw}. The official project page reports 2,540,044 full records, 49 features, and nine attack categories; the mirror contains 480,629 fewer rows, and its row-selection and filtering lineage remain unresolved.

The development predicate is \nolinkurl{ltime < 1424219008}, admitting 1,023,197 rows. The remaining frozen pilot segment contains 1,036,218 rows. The development roles were assigned using the manifest's label-blind hash rule over 60-raw-tick blocks; they interleave in time. This is offline development, not train-before-validation forecasting. Earlier unlabeled development events can provide retained context across those roles.

\begin{table*}[t]
\caption*{Frozen pilot role support}
\centering
\small
\setlength{\tabcolsep}{4pt}
\begin{tabularx}{\textwidth}{Xrr}
\toprule
Frozen pilot role & Rows & Reconnaissance rows \\
\midrule
Training & 775,008 & 1,361 \\
Tuning & 104,620 & 215 \\
Calibration & 143,569 & 164 \\
Final test & 1,036,218 & 11,617 \\
\bottomrule
\end{tabularx}
\end{table*}

These are previously audited aggregate supports. The retention follow-up did not read labels or reconstruct these counts from labels.

\subsection{Observation and state semantics}

Raw \nolinkurl{ltime} values define a \textbf{proxy release order}, not verified physical arrival or completed-flow export times. Completed-flow fields are treated as available under this convention. The 2,366 duration-inconsistent records identified in the pilot documentation were retained and flagged without clock repair. Raw ticks are not converted into claims about physical response time.

For the pilot, retained capacity is \nolinkurl{N = 256} records and the horizon is \nolinkurl{W = 300} raw ticks. Before querying a group at tick \nolinkurl{t}, records with retained tick strictly less than \nolinkurl{t - W} expire. Thus the lower horizon boundary is inclusive, subject to earlier capacity eviction. All records with the same exact \nolinkurl{ltime} query the same expired prior-group state. They cannot supply history to another record in that group. Only after feature extraction does the group commit in the frozen order:

\begin{quote}\small
\nolinkurl{(source-address bytes, destination-address bytes,}\par
\nolinkurl{ protocol-group code, destination port, flow identifier)}\par
\end{quote}

Endpoint identifiers are used internally for state keys and deterministic ordering. No raw addresses or row-level labels are reproduced here. The source adapter maps its 49 observed, case-sensitive protocol literals to TCP, UDP, or other under a source-specific mapping.

The service key \nolinkurl{v} is the ordered pair \nolinkurl{(protocol-group code, destination port)}. Thus this canonical order refines the theoretical \nolinkurl{(s,d,v)} order with a flow-identifier tie-break. Within an equal-tick group, tied \nolinkurl{(s,d,v)} updates have identical retained endpoint/\allowbreak{}service tuples and expiration ticks, so this tie-break orders record identities without changing their projected endpoint/\allowbreak{}service updates. This crosswalk does not extend the theoretical FIFO ring's byte-memory or cost guarantees to the Python source-aware policies.

The pilot resets retained state before the final segment. Retained-state capacity is separate from input batches, same-tick buffers, feature matrices, models, label storage, and other offline workspace. A 256-record cap is not a fixed-total-memory or fixed-computation guarantee.

\subsection{External-corpus feasibility}

The UWF-ZeekData feasibility audit did not qualify an independent corpus for the intended contemporaneous benign-versus-reconnaissance, multi-scanner experiment. The audited Fall22 benign and reconnaissance populations were separated by capture period. The apparently mixed 2022 CSVs also separated benign and attack traffic by period, contained substantial duplication, and exhibited a class-associated timestamp inconsistency. Multi-scanner attack parquet files lacked benign background traffic.

Deduplication alone does not remove these period and population limitations. The rejection applies to the intended experiment; it is not a statement that the artifacts have no possible research use. No UWF predictive result enters this report, and no alternative independent corpus is currently qualified.

\section{Frozen Predictive Pilot}

\subsection{Models and decision policy}

The pilot compared three frozen \nolinkurl{HistGradientBoostingClassifier} models without hyperparameter search:

\begin{itemize}
\item \textbf{B0:} the frozen current-flow feature allowlist.
\item \textbf{B1:} B0 inputs plus source event count, distinct destination-port count, and retained occupancy.
\item \textbf{B2:} B1 inputs plus five retained relational quantities: distinct destinations per source (\ensuremath{D_s}), distinct sources per destination (\ensuremath{I_d}), service-expanded source out-degree (\ensuremath{O_s}), exact source--destination--service event multiplicity (\ensuremath{R_{sdv}}), and distinct services per source--destination (\ensuremath{V_{sd}}).

\end{itemize}
The primary predictive contrast was B2 minus B1. No IP, flow identifier, timestamp, label, split role, hash, or \nolinkurl{ct_} field was a learned feature. The lexical \nolinkurl{state} vocabulary was learned once from training flows; unseen or absent states became missing. Numeric nonfinite values likewise became missing without dropping rows.

Each classifier used training labels only. A separate nonnegative-slope Platt map was fitted to calibration outcomes using the frozen regularized, constrained objective and deterministic solver. This procedure permits a constant solution and does not itself guarantee calibration.

Each raw, calibrated, and reference score stream received its own threshold from tuning-normal scores. For \nolinkurl{n} such scores, the rule allowed at most \nolinkurl{floor(n/100)} alerts and selected the smallest observed score level satisfying that constraint, or positive infinity if needed. Alerts used \nolinkurl{score >= threshold}; ties were indivisible. The 1\% empirical tuning constraint was not a population or final-test false-positive-rate guarantee.

A constant predictor based on calibration reconnaissance frequency was included. Models, calibration maps, thresholds, and the constant reference were frozen before final-label access in the completed repeat.

\subsection{Test-exposure and check scope}

The completed pilot, run \nolinkurl{20261023T132603-8c37fa79}, was a \textbf{disclosed repeat} after an interrupted attempt with possible final-label exposure and no recoverable fitted checkpoints. It was not a same-model continuation. Final-test label metadata had also been inspected during earlier data auditing. Consequently, the final segment cannot be described as a pristine test or as supporting independent dataset selection.

The scientific choices remained frozen for the completed repeat. Its fitted artifacts were committed at approximately 14:10:52 UTC, before the recorded first final-label access at 14:11:50 UTC. This establishes the recorded ordering within that run; it does not erase earlier exposure.

An internal check covered scoped reporting with limitations after checking provenance, component hashes, support and metric consistency, and resampling configuration. Whole-test numerical reproduction was not established because the final score arrays and resampling draws were not retained. The results below are the previously checked frozen-pilot outputs, not newly recomputed estimates.

\subsection{Calibrated operating-point results}

Table 1 reproduces the reported six-decimal values from the pilot draft. Average precision is the documented step-wise PR-AUC estimator.

\begin{table*}[t]
\caption{Frozen calibrated pilot results. Each stream uses its own previously frozen operating threshold.}
\centering
\small
\setlength{\tabcolsep}{4pt}
\begin{tabularx}{\textwidth}{Xrrrrrrr}
\toprule
Model & \shortstack{Recon.\\recall} & \shortstack{Benign\\FPR} & \shortstack{Other-attack\\misattribution} & AUROC & \shortstack{Average\\precision} & \shortstack{Brier\\score} & \shortstack{Log\\loss} \\
\midrule
B0 & 0.478092 & 0.063653 & 0.303534 & 0.734002 & 0.137139 & 0.014194 & 0.156281 \\
B1 & 0.478437 & 0.066787 & 0.303317 & 0.735021 & 0.123377 & 0.013746 & 0.189880 \\
B2 & 0.436602 & 0.036291 & 0.231722 & 0.678465 & 0.120835 & 0.014084 & 0.125154 \\
Calibration-prior constant & 0 & 0 & 0 & 0.500000 & 0.011211 & 0.011187 & 0.077081 \\
\bottomrule
\end{tabularx}
\end{table*}

B2 had lower recall than B1 by 4.18 percentage points and lower benign FPR by 3.05 percentage points, as rounded in the pilot report. These are different operating points, not a matched-final-FPR comparison. B2 also had lower AUROC and average precision, higher Brier score, and lower log loss and other-attack misattribution. Thus the added relational features did not establish a general predictive advantage.

The constant reference had lower Brier score and log loss than the learned models while detecting no reconnaissance. Its inclusion is important in this low-prevalence setting. All learned models exceeded the 1\% empirical tuning FPR target on the final segment; no thresholds were retuned afterward.

The pilot used 500 paired resampling replicates of nonempty 300-tick final-segment blocks, with PCG64 seed 2026 and fixed fitted components. Its percentile summaries describe within-capture variability only. They do not establish population coverage, independent episodes, or fitting uncertainty. No inferential conclusion is drawn here from those ranges.

\section{Development Mechanism Audit}

After the mixed pilot result, a label-blind development-only audit compared retained history \nolinkurl{R} with the full prior 300-tick reference window \nolinkurl{Q}. The audit found that capacity eviction, rather than horizon expiration, governed the FIFO state on this trace:

\begin{itemize}
\item Mean occupancy was 255.93 records, with median 256.
\item There were 1,022,685 capacity victims among 1,023,197 insertions.
\item Capacity eviction occurred in 44,350 of 44,390 groups.
\item There were 256 horizon removals, at the sparse terminal tail, and 256 terminal censored survivors.
\item Completed record lifetimes had median 11 ticks and 95th percentile 13 ticks.

\end{itemize}
The audit's pooled equal-source retained/\allowbreak{}reference ratios were reported as 0.1693 for \ensuremath{O_s} and 0.1989 for \ensuremath{V_{sd}}. These indicated substantial loss of the corresponding reference history under the specified state. They did not establish that the missing history was predictive or that its loss accounted for the B2 result.

The audit nominated a source-stratified allocation hypothesis for a later protocol. An appended erratum corrected \ensuremath{R_{sdv}}: it is \textbf{event multiplicity for the exact source--destination--service relation}, not the number of distinct sources per destination--service. This draft uses the corrected meaning. The audit and follow-up also have different measurement inclusion rules: the follow-up excludes block 0 from measurement. Their pooled values should therefore not be treated as measurements over identical query populations.

The audit's optional development-score component was unavailable because no provenance-qualified designated development score artifact was found. No predictive linkage between retention measurements and development scores was evaluated.

\section{Prespecified Retention-Allocation Study}

\subsection{Scope and measurement period}

The \nolinkurl{retention-mechanism-v1} protocol was specified and internally checked before its single execution, but after the pilot and mechanism audit. It is a computational follow-up on reused development data.

The runner streamed the development segment without globally sorting it, using Parquet input batches of at most 65,536 rows and a hard 256-record same-tick group buffer. The observed maximum development group contained 140 records. A decrease in source \nolinkurl{ltime} or violation of the declared bounds was a correctness failure.

The follow-up used the same inclusive 300-tick horizon and prior-group query semantics across all arms. Its blocks were defined by:

\[
b(t)=\left\lfloor\frac{t-1421927414}{300}\right\rfloor.
\]

Block 0 updated state but contributed no measurements. Measurement covered the fixed index range 1 through 151, including the partial final block observed at \nolinkurl{ltime <= 1421972724}. The fixed subblocks were:

\begin{itemize}
\item \textbf{S1:} blocks 1--50;
\item \textbf{S2:} blocks 51--100;
\item \textbf{S3:} blocks 101--151.

\end{itemize}
These are measurement ranges within one continuous replay, not independent trials. State was not reset at the subblock boundaries.

No labels, attack categories, models, scores, calibration maps, thresholds, or checkpoints were inputs to this study. The reader stopped at the first row at or beyond the development boundary, observed only its \nolinkurl{ltime} for closure, and discarded it before extracting endpoint, protocol, or label fields. No final-segment record was admitted or evaluated. Whole-file source hashing and the boundary-tick observation are provenance operations, not final-test performance evaluation.

\subsection{Reference window and five arms}

The reference \nolinkurl{Q} contained all events from strictly earlier groups within the 300-tick horizon, without a capacity cap. It existed to define measurement denominators. Retention policies could not consult it, and it supplied no classifier inputs.

The five arms were fixed before execution:

\begin{table*}[t]
\caption*{Prespecified retention arms}
\centering
\small
\setlength{\tabcolsep}{4pt}
\begin{tabularx}{\textwidth}{lrX}
\toprule
Arm & Record cap & Eviction policy \\
\midrule
FIFO-64 & 64 & Global insertion-order FIFO \\
FIFO-256 & 256 & Global insertion-order FIFO; primary baseline \\
FIFO-1024 & 1024 & Global insertion-order FIFO \\
SRC-256 & 256 & Evict the oldest record of the currently most occupied retained source \\
DIV-256 & 256 & Source-balanced victim-source selection, with preference for evicting a duplicate relation \\
\bottomrule
\end{tabularx}
\end{table*}

\textbf{FIFO} removes the oldest live retained record when a full state must admit an incoming record. \textbf{SRC} instead chooses the source with the largest live retained occupancy, breaking source ties by lexicographically smallest address-byte string, and removes that source's oldest live record. SRC uses current retained-state metadata; it does not consult a global source census or known scanner identities. It does not guarantee equal occupancy for all sources.

FIFO and SRC evict before admitting when full, so the incoming record cannot be its own eviction victim. \textbf{DIV} deliberately admits first. It then uses the same largest-occupancy source selection and scans that source's records from oldest to newest, evicting the first whose \nolinkurl{(destination, protocol group, destination port)} relation has another live occurrence from the same source. If none exists, it evicts the oldest record of that source. This post-admission rule makes the incoming relation visible to the duplicate check. The cap is respected after each completed admission-and-eviction operation; admission semantics are not identical across all arms.

All policies used only allowed retained endpoint, protocol, port, tick, and canonical-order metadata. The primary comparison was SRC-256 versus FIFO-256. The other arms were descriptive and could not replace that comparison after outcomes were observed.

\subsection{Endpoint definitions}

For a query event \nolinkurl{e} with source \nolinkurl{s} and destination \nolinkurl{d}, define:

\begin{itemize}
\item \ensuremath{O_s(e)}: the number of distinct \nolinkurl{(destination, protocol group, destination port)} relations represented by records from \nolinkurl{s} in the applicable history.
\item \ensuremath{V_{sd}(e)}: the number of distinct \nolinkurl{(protocol group, destination port)} services represented for \nolinkurl{(s,d)} in that history.

\end{itemize}
For either quantity \nolinkurl{X}, a query is eligible only when its reference value \ensuremath{X_Q(e)} is positive. Its retention ratio for arm \nolinkurl{a} is:

\[
r_{a,X}(e)=\frac{X_{R_a}(e)}{X_Q(e)}.
\]

For measurement region \nolinkurl{I}, let \ensuremath{E_{s,I,X}} be the eligible queries from source \nolinkurl{s}, and let \ensuremath{S_{I,X}} be the sources with at least one such query. The equal-source mean is:

\[
\overline r_{a,I,X}
=\frac{1}{|S_{I,X}|}
\sum_{s\in S_{I,X}}
\left[\frac{1}{|E_{s,I,X}|}
\sum_{e\in E_{s,I,X}}r_{a,X}(e)\right].
\]

The endpoint is:

\[
M_{a,I}=\frac{1}{2}
\left(\overline r_{a,I,O_s}+\overline r_{a,I,V_{sd}}\right).
\]

The pooled endpoint applies this definition to all eligible measured queries in S1, S2, and S3 together. It is not the simple average of the three subblock endpoints. Component source sets can differ. Defined zero ratios are included; zero reference denominators are excluded, not replaced by zero.

\subsection{Fixed decision rule and secondary measurements}

The primary contrast was:

\[
\Delta M=M_{\text{SRC-256}}-M_{\text{FIFO-256}}.
\]

The prespecified rule required pooled \ensuremath{\Delta M\geq0.05} and positive \ensuremath{\Delta M} in at least two of S1, S2, and S3. Required component means had to be defined with at least two contributing sources. Otherwise, the decision criterion was unavailable. This was an engineering materiality and descriptive stability rule, not an inferential test or a reconnaissance-specific criterion.

Secondary measurements included DIV-256 versus FIFO-256, the FIFO capacity arms, event-weighted ratios, and the retained/\allowbreak{}reference exact-relation event-multiplicity ratio \ensuremath{R_{sdv}}. Event-weighted ratios average eligible query ratios directly, rather than first averaging within sources and assigning each source equal weight. Neither secondary comparisons nor alternative weighting could override the declared primary rule.

\section{Retention Results}

\subsection{Pooled endpoints and primary contrast}

\begin{table*}[t]
\caption{Pooled retention endpoints for all prespecified arms. Values are rounded to six decimals from \nolinkurl{summary.json}; \ensuremath{O_s} and \ensuremath{V_{sd}} columns are equal-source mean ratios. Arithmetic and the decision rule use the stored machine-precision values, not the rounded display.}
\centering
\small
\setlength{\tabcolsep}{4pt}
\begin{tabularx}{\textwidth}{Xrrr}
\toprule
Arm & Pooled \ensuremath{M} & \shortstack{Equal-source\\\ensuremath{O_s}} & \shortstack{Equal-source\\\ensuremath{V_{sd}}} \\
\midrule
FIFO-64 & 0.091471 & 0.073888 & 0.109055 \\
FIFO-256 & 0.183633 & 0.168766 & 0.198501 \\
FIFO-1024 & 0.319540 & 0.318420 & 0.320660 \\
SRC-256 & 0.624371 & 0.621398 & 0.627345 \\
DIV-256 & 0.645849 & 0.646344 & 0.645355 \\
\bottomrule
\end{tabularx}
\end{table*}

\begin{table*}[t]
\caption{Prespecified primary contrast and fixed subblocks. Source counts are eligible contributing sources for \ensuremath{O_s} /\allowbreak{} \ensuremath{V_{sd}}, not scanner counts.}
\centering
\small
\setlength{\tabcolsep}{4pt}
\begin{tabularx}{\textwidth}{Xrrrr}
\toprule
Aggregation & FIFO-256 \ensuremath{M} & SRC-256 \ensuremath{M} & \shortstack{SRC-256 minus\\FIFO-256} & Sources \ensuremath{O_s} /\allowbreak{} \ensuremath{V_{sd}} \\
\midrule
Pooled & 0.183633 & 0.624371 & 0.440738 & 35 /\allowbreak{} 34 \\
S1 & 0.177710 & 0.622169 & 0.444459 & 35 /\allowbreak{} 34 \\
S2 & 0.083780 & 0.619761 & 0.535980 & 25 /\allowbreak{} 25 \\
S3 & 0.136080 & 0.622703 & 0.486623 & 25 /\allowbreak{} 25 \\
\bottomrule
\end{tabularx}
\end{table*}

The pooled contrast exceeded 0.05 and all three subblock contrasts were positive. Required support was available, so the frozen rule was satisfied. The subblocks are dependent descriptions of the same trace, not three replications.

DIV-256's pooled difference from FIFO-256 was 0.462216. It remains a secondary descriptive comparison. Its value does not promote DIV to the primary policy, justify a predictive claim, or authorize further selection.

\subsection{Event-weighted qualification}

\begin{table*}[t]
\caption{Pooled event-weighted ratios for the primary arms. These are averages over eligible queries, using the same reference-positive eligibility rule.}
\centering
\small
\setlength{\tabcolsep}{4pt}
\begin{tabularx}{\textwidth}{Xrr}
\toprule
Quantity & FIFO-256 & SRC-256 \\
\midrule
\ensuremath{O_s} & 0.066887 & 0.061185 \\
\ensuremath{V_{sd}} & 0.072118 & 0.065100 \\
\bottomrule
\end{tabularx}
\end{table*}

Both event-weighted ratios were lower for SRC-256. This distinction is central to interpretation. Equal-source averaging gives each eligible source the same weight regardless of how many eligible queries it contributes; event weighting assigns more aggregate weight to sources with more such queries. The observed directions differ under these two objectives. The positive primary result therefore supports higher equal-source retention ratios, not uniformly improved retention or better coverage of the query stream under every weighting.

The repetition diagnostic shows the same need to specify weighting. Pooled equal-source \ensuremath{R_{sdv}} was 0.143288 for FIFO-256 and 0.412612 for SRC-256, whereas its event-weighted values were 0.068044 and 0.066753, respectively. These are exact-relation event-multiplicity ratios, not distinct-source coverage or prediction metrics.

\subsection{Measurement support}

The study measured 1,015,598 queries after 7,599 warm-up queries. S1, S2, and S3 contributed 343,995, 329,101, and 342,502 measured queries, respectively. For pooled \ensuremath{O_s}, 1,015,025 queries were eligible and 573 had zero reference denominator. For \ensuremath{V_{sd}}, 1,013,333 were eligible and 2,265 had zero reference denominator.

Eligible per-source query support was uneven. Pooled \ensuremath{O_s} source support ranged from 1 to 101,046 queries, with median 712.0; \ensuremath{V_{sd}} support ranged from 2 to 101,034, with median 1029.0. Nevertheless, each contributing source received equal weight within the corresponding component mean. A source with very sparse support can therefore contribute as much to that mean as a frequently observed source. The reported endpoint is valid under its declared definition, but its magnitude should not be interpreted as the experience of a typical event or a typical scanner.

No confidence interval, p-value, or inferential conclusion was computed for the retention comparison.

\section{Execution and Internal-Check Provenance}

The retention study completed one authorized supervised run. The execution receipts record matching expected/\allowbreak{}before/\allowbreak{}after source SHA-256 values and matching before/\allowbreak{}after file identities; this does not resolve the mirror's curation lineage. The reader observed 1,023,198 physical rows, admitted 1,023,197 development rows, and discarded the boundary record after reading its tick. It completed 44,390 groups with maximum group size 140.

The worker was pinned to one CPU. Runner and supervisor wall times were 164.144 and 164.423 seconds, respectively. Peak recorded RSS was 189,206,528 bytes, approximately 189 MB in decimal units. The final output tree was 97,979 bytes after the supervisor receipt; the runner's accounting checkpoint recorded 97,279 bytes before that receipt. The supervisor returned code 0 without timeout, within the declared 2-hour, 8-GiB RSS, and 2-GiB output budgets. These are measurements of the combined offline replay and measurement job, not per-arm costs or online latency.

The pre-execution internal check covered the protocol/\allowbreak{}implementation package, its frozen dependency closure, endpoint semantics, isolation rules, and execution gates. The subsequent read-only result check examined output hashes, protocol/\allowbreak{}manifest/\allowbreak{}freeze/\allowbreak{}authorization links, completion and resource evidence, required support, and primary/\allowbreak{}subblock arithmetic. It did not replay the real source or provide external replication. The recorded internal outcome was \nolinkurl{APPROVE_RESULT} for the bounded result statement.

AI agents (Astra and Terra) were used for internal checks of implementation, recorded provenance, consistency, and arithmetic within documented scopes. The retention-result checks examined recorded outputs and decision-rule arithmetic without replaying the source; whole-test numerical reproduction of the predictive pilot was not established. These checks were neither scholarly peer review nor external replication. The author is responsible for the manuscript and its claims.

The earlier predictive pilot has a different check scope: its provenance and internal result consistency were checked with limitations, while whole-test numerical reproduction remained unestablished. These outcomes should not be conflated.

\section{Discussion}

\subsection{What the evidence establishes}

The audit documented a capacity-dominated FIFO state on the development trace. The follow-up then measured a different retention allocation under a shared record cap, horizon, event stream, and reference definition. The equal-source endpoint was higher for the prespecified SRC policy, while the corresponding event-weighted ratios were lower. This combination makes the weighting choice part of the substantive result.

The report thus connects a mixed predictive finding to a narrower measurement question about retained history. It does not turn the predictive finding into a success. The original B2 model was not refitted or evaluated with SRC retention, and there is no measured relationship showing that the higher equal-source ratios would improve recall, false-positive rate, score ranking, or probability estimates.

\subsection{What remains unresolved}

Reference history is a computational comparator, not evidence that every reference relation is relevant to reconnaissance. More retained diversity can coexist with less event-weighted retention, and neither quantity alone determines predictive usefulness. The recorded aggregates also do not identify which scanners, if any, would benefit.

The capacity audit does not isolate the reason for B2's predictive result. Feature definitions, class prevalence, model behavior, and other aspects of the frozen design remain possible contributors. This study does not distinguish among those explanations.

Any future predictive comparison would be a separate study requiring a defensible population, an outcome-untouched evaluation design, and an internally checked protocol with fixed fitting and decision rules. The UWF audit supplies a documented feasibility rejection, not a substitute evaluation. This manuscript does not depend on an additional experiment to report its existing bounded findings.

\section{Limitations}

\begin{enumerate}
\item \textbf{Reused trace and post-pilot motivation.} The pilot, audit, and allocation study share development data. Specifying the allocation protocol before its own run does not undo the earlier exposure or make the follow-up independent. Resplitting this trace would not create independent confirmation.

\item \textbf{Reconnaissance concentration and limited scanner diversity.} Prior feasibility work found all 1,740 development reconnaissance rows in 300-tick blocks 0--23, with four recorded scanner sources. Blocks 24--151 had no development reconnaissance. Consequently, S2/\allowbreak{}S3 positivity cannot be interpreted as evidence that the mechanism preserves reconnaissance history across later periods. The follow-up's 35/\allowbreak{}34 pooled eligible sources are all-source measurement supports, not 35/\allowbreak{}34 scanners. A conventional chronological predictive split was rejected because the proposed later roles would lack positives; no source-disjoint predictive study was performed.

\item \textbf{Sparse and unequal support.} Some sources contribute very few eligible queries but receive equal weight. Reference-zero queries are excluded by definition, and component source sets differ. The large pooled endpoint difference does not characterize every source, every event, or every relation. Event-weighted declines directly limit broader retention claims.

\item \textbf{No established predictive relationship.} Retained/\allowbreak{}reference ratios have not been shown to predict detection utility. No SRC-based classifier, calibrated probability model, threshold policy, or predictive comparison was fitted in the retention study. Labels and scores were absent from that experiment; the earlier audit's development-score component was unavailable.

\item \textbf{Pilot test exposure and incomplete numerical reproduction.} The pilot final segment had prior label-metadata exposure, and an interrupted attempt may have accessed its labels. The completed run was a disclosed repeat with unchanged scientific choices. Its within-run checkpoint ordering does not restore pristine evaluation. Full independent reproduction of its scores and resampling calculations was not established.

\item \textbf{Record budgets are not equal resource budgets.} SRC and FIFO share a 256-record cap, but policy metadata, victim-selection work, and transient workspace can differ. DIV also uses post-admission eviction. No equal-byte, equal-compute, per-arm timing, or cost-optimality conclusion follows. The uncapped reference and offline workspace are outside the retained-state cap.

\item \textbf{Proxy timing and source curation.} Raw \nolinkurl{ltime} order is an artificial release convention. Physical availability of completed-flow fields, export timing, and duration inconsistencies remain unresolved. The mirror's missing-record and filtering mechanisms are also unresolved. The measured ticks and job duration do not establish physical detection delay.

\item \textbf{No qualified external evaluation.} UWF did not satisfy the intended contemporaneous benign/\allowbreak{}reconnaissance, multi-scanner requirements after its duplication and timing issues were examined. Using other attack tactics as negatives would change the task and would not measure benign false positives. No alternative independent corpus has been qualified.

\item \textbf{Dependent descriptions and fixed scope.} The three subblocks share one trace and continuous state. Their signs implement the declared descriptive rule; they do not provide independent trials or inferential evidence. The 0.05 margin is an engineering criterion, not a utility threshold. Secondary DIV and FIFO-capacity comparisons cannot broaden the primary result.

\item \textbf{Internal checks are bounded verification.} AI-agent internal checks support the documented computational and reporting checks. They are not external replication or evidence about an unseen population. The retention study makes no final-test performance claim; only the earlier frozen pilot supplies the historical predictive results summarized here.

\end{enumerate}
\section{Conclusion}

On this already-used UNSW development trace, source-balanced 256-record retention preserved more reference \ensuremath{O_s}/\allowbreak{}\ensuremath{V_{sd}} history than global FIFO-256 under the prespecified equal-source endpoint and stability rule.

\section{Reproducibility and Artifact Map}

Paths below are relative to the repository root \nolinkurl{/home/jsantillana/projects/anovel-paper} and are printed as local artifact identifiers. They identify existing evidence, not a newly executed analysis.

\sloppy
\begin{description}
\item \textbf{Frozen-pilot methods, reported metrics, and exposure limitations}\par \nolinkurl{manuscript/PILOT_DRAFT.md}
\item \textbf{Original completed predictive-pilot result record}\par \nolinkurl{results/frozen_repeat_v1/artifacts/pilot_experiment_20261023T132603-8c37fa79.json}
\item \textbf{Pilot result-check scope and limitations}\par \nolinkurl{validation/results_round1/REVIEW.md}
\item \textbf{Capacity audit, nominated mechanism, and \ensuremath{R_{sdv}} erratum}\par \nolinkurl{data_audits/development_mechanisms_v1/DEVELOPMENT_MECHANISM_FINDINGS.md}
\item \textbf{Reviewed bounded retention-result statement}\par \nolinkurl{data_audits/development_mechanisms_v1/RETENTION_MECHANISM_RESULT.md}
\item \textbf{Follow-up semantics, endpoint, rule, restrictions, and gates}\par \nolinkurl{RETENTION_MECHANISM_PROTOCOL.md}
\item \textbf{Frozen retention-study manifest}\par \nolinkurl{configs/retention_mechanism_v1.json}
\item \textbf{Full stored endpoints, weightings, supports, and provenance}\par \nolinkurl{results/retention_mechanism_v1/summary.json}
\item \textbf{Human-readable retention-run report}\par \nolinkurl{results/retention_mechanism_v1/REPORT.md}
\item \textbf{Reviewed 26-file dependency freeze}\par \nolinkurl{validation/retention_mechanism_v1/FREEZE.json}
\item \textbf{Pre-execution internal check}\par \nolinkurl{validation/retention_mechanism_v1/REVIEW.md}
\item \textbf{Read-only retention result check and reported output digests}\par \nolinkurl{validation/retention_mechanism_v1/RESULT_REVIEW.md}
\item \textbf{External-corpus feasibility rejection}\par \nolinkurl{data_audits/uwf_zeek_feasibility_v1/AUDIT.md}
\item \textbf{Verified bibliography and mirror-lineage record}\par \nolinkurl{literature/VERIFIED_REFERENCES.md}
\item \textbf{Chronology, authorizations, internal checks, and temporal-support finding}\par \nolinkurl{CONTINUE.md}
\end{description}
\fussy

Selected recorded SHA-256 digests:

\begin{quote}\small
\nolinkurl{Retention protocol:}\par
\nolinkurl{7c5ead5c15cb6cf590846407055f3a5e65345324e47ea6b41256d2edbb828d6c}\par
\nolinkurl{Retention manifest:}\par
\nolinkurl{dfc0d1101096247d28808070212e35c967e9e41f519f4620513bfe5a2301d10d}\par
\nolinkurl{Retention dependency freeze:}\par
\nolinkurl{3a98848112dd586092b8241eaf3183e7ac67fe508cdfa9b7dfed190e7fe989d4}\par
\nolinkurl{Retention summary.json (as recorded in the result review):}\par
\nolinkurl{7842e75ab92fb15d304c660f3654f3b3743fb61aa92d83feeb4f9e2d6942a0a0}\par
\end{quote}

Table 1 retains the reporting precision of the pilot draft. Tables 2--4 round the stored retention values to six decimals; \nolinkurl{summary.json} retains the machine-precision results. Its \nolinkurl{block_count} field counts distinct block indices containing observed groups (\nolinkurl{len(block_counts)} in \nolinkurl{retention/replay.py}), not the length of the configured range. Hence \nolinkurl{block_count=150} with \nolinkurl{block_min=0} and \nolinkurl{block_max=151} denotes 150 nonempty indices within 152 possible indices, including warm-up; it does not change the fixed measurement/\allowbreak{}subblock ranges or imply a reset or omitted input row. No new source pass was performed to identify the empty indices. Aggregate scanner-support statements come from the earlier feasibility record, not from a new label read. Reproduction claims should distinguish the available code, protocol, and recorded outputs from independently rerunning the original experiments.

\section*{Citation and Lineage Verification}

The cited works' bibliographic metadata was verified against Crossref and publisher or project records; their attributed scope is limited to the statements in Section~\ref{sec:related}. The dataset mirror's row-selection lineage remains unresolved. Local artifact links above are evidence pointers, not substitutes for scholarly references.

\bibliography{references}

\pdfinfo{/Author (Juan S. Santillana)}

\end{document}